\RequirePackage[l2tabu, orthodox]{nag}
\RequirePackage{snapshot}

\documentclass[9pt,onecolumn]{extarticle}
\makeatletter
\if@twocolumn
  \usepackage[dvips,letterpaper,top=0.5in, bottom=0.5in, left=0.75in, right=0.5in,includefoot,heightrounded]{geometry}
\else
  \usepackage[dvips,letterpaper,margin=1in,includefoot,heightrounded]{geometry}
\fi

\usepackage{srcltx}

\usepackage{amsmath}
\usepackage{amssymb,amsfonts}

\usepackage{abstract}

\usepackage{graphicx}
\usepackage[usenames,dvipsnames]{color}
\usepackage[usenames,dvipsnames]{xcolor}
\usepackage{subfigure}

\usepackage{booktabs}

\usepackage{setspace}
\usepackage{flushend}

\usepackage{multicol}

\usepackage{mwe}

\usepackage{url}
\usepackage{bm}

\usepackage{multirow}

\usepackage[short,12hr]{datetime}

\usepackage{hyperref}
\usepackage[hyperpageref]{backref}

\makeatletter

\newenvironment{rsmallmatrix}{\null\,\vcenter\bgroup
  \Let@\restore@math@cr\default@tag
  \baselineskip6\ex@ \lineskip1.5\ex@ \lineskiplimit\lineskip
  \ialign\bgroup\hfil$\m@th\scriptstyle##$&&\thickspace\hfil
  $\m@th\scriptstyle##$\crcr
}{%
  \crcr\egroup\egroup\,%
}

\newcommand{\printtitle}{%
\makeatletter
\if@twocolumn

\twocolumn[%
  \maketitle
  \begin{onecolabstract}
    \myabstract
  \end{onecolabstract}
  \begin{center}
    \small
    \textbf{Keywords}
    \\\medskip
    \mykeywords
  \end{center}
  \bigskip
]
\saythanks
\else
  \maketitle
  \begin{onecolabstract}
    \myabstract
  \end{onecolabstract}
  \begin{center}
    \small
    \textbf{Keywords}
    \\\medskip
    \mykeywords
  \end{center}
  \bigskip
  \onehalfspacing
\fi
\makeatother
}

\title{%
Efficient Computation of the 8-point DCT via Summation by Parts}

\author{%
D.~F.~G.~Coelho%
\thanks{%
D. F. G. Coelho is with the
Department of Electrical and Computer Engineering,
University of Calgary, Calgary, Canada.}
\quad
R.~J.~Cintra%
\thanks{%
R. J. Cintra is with the
Signal Processing Group,
Departamento de Estat\'istica,
Universidade Federal de Pernambuco
and
the
Department of Electrical and Computer Engineering,
University of Calgary, Calgary, Canada.
E-mail: \protect\url{rjdsc@de.ufpe.br}}
\quad
V.~S.~Dimitrov%
\thanks{%
V. S. Dimitrov is with the
Department of Electrical and Computer Engineering,
University of Calgary, Calgary, Canada.}
}

\date{}

\newcommand{\myabstract}{%
This paper introduces
a new fast algorithm for the
\mbox{8-point} discrete cosine transform (DCT) based on
the summation-by-parts formula.
The proposed method
converts the DCT matrix into an alternative
transformation matrix
that can be decomposed
into sparse matrices
of
low multiplicative complexity.
The method is capable
of
scaled
and exact
DCT computation and its associated fast algorithm
achieves the theoretical minimal multiplicative complexity
for the 8-point DCT.
Depending on the nature of the input signal
simplifications can be introduced and the
overall complexity of the proposed algorithm
can be further reduced.
Several types of input signal are analyzed:
arbitrary, null mean, accumulated, and
null mean/accumulated signal.
The proposed tool
has potential application in
harmonic detection,
image enhancement, and feature extraction,
where input signal DC level
is discarded and/or the signal is required to be integrated.
}

\newcommand{\mykeywords}{%
DCT,
Fast Algorithms,
Image Processing
}

\begin{document}

\printtitle

\section{Introduction}

Discrete transforms play a central role in signal processing.
Noteworthy methods include
trigonometric transforms---such as the
discrete Fourier transform (DFT)~\cite{Oppenheim2009},
discrete Hartley transform (DHT)~\cite{Oppenheim2009},
discrete cosine transform (DCT)~\cite{Britanak2007},
and
discrete sine transform (DST)~\cite{Britanak2007}---as well
as the
Haar and Walsh-Hadamard transforms~\cite{Ahmed1975}.
Among these methods,
the DCT
has been applied in several practical contexts:
noise reduction~\cite{Gupta2012},
watermarking methods~\cite{An2009},
image/video compression techniques~\cite{Britanak2007},
and
harmonic detection~\cite{Britanak2007},
to cite a few.
In fact,
when processing
signals modeled   as a   stationary Markov-1 type random process,
the
DCT behaves as the asymptotic case of
the optimal Karhunen--Lo\`eve transform
in terms of data decorrelation~\cite{Britanak2007}.
This approximation holds true when
the
correlation coefficient of the related stochastic process tends to the unit,
which is the case for many real signals---specially images~\cite{Britanak2007}.
Moreover,
the recent increase in image/video processing demand
for consumer electronics~\cite{Chien2013}
and big data manipulation~\cite{Wigan2013}
emphasizes the necessity for
fast and efficient DCT computation~\cite{Ji2002}.

As a consequence,
the 8-point DCT
is adopted in several
image and video coding schemes~\cite{Bhaskaran1995},
such as
JPEG~\cite{Wallace1992},
\mbox{MPEG-1}~\cite{Wallace1992},
\mbox{H.264}~\cite{Wiegand2003},
HEVC~\cite{Pourazad2012},
AVS China~\cite[p.~61]{Rao2014}, and VP-10~\cite[p.~165]{Rao2014}.
Aiming at minimizing the computational cost of
the  DCT evaluation,
a
number of
fast algorithms for the 8-point DCT have been proposed,
including
Chen's DCT algorithm~\cite{Wen-HsiengChen2003},
Lee method~\cite{Lee1984},
Loeffler algorithm~\cite{Loeffler1989},
Feig-Winograd DCT factorization~\cite{Feig1992},
and
the
Arai DCT~\cite{Arai1988}.

Multiplication operations
as required by DCT and others discrete-time transforms
can be
implemented via long sequences of additions,
bit-shifting operation, and sign changes~\cite{Hamming1989}.
Thus,
algorithms that require multiplications
often
have
higher computational costs~\cite{Blahut2010}.
Therefore,
above-mentioned methods
were developed
in order to reduce the overall
number of multiplications~\cite{Britanak2007}.
The Arai DCT is particularly
useful because it furnishes a scaled version of the DCT spectrum.
In some applications
such as harmonic detection~\cite{Limin2007,Zheng2010}
and JPEG-like image compression~\cite{Wallace1992,Bhaskaran1995},
the scaled DCT is often a sufficient tool.
This is because in these contexts
only the relative value of the spectrum is necessary.
Therefore,
part of the cost of computing the DCT can be avoided~\cite{Britanak2007}.

Among the fundamental mathematical tools,
we  separate the summation-by-parts technique~\cite[p.~54]{Graham1989},
which is the discrete-time counterpart
for the well-known integration-by-parts method~\cite[p.~144]{Apostol1981}.
Although
applied in several contexts
such as
computational physics for approximate second derivatives~\cite{Mattssona2004},
approximations of the linear advection-diffusion equation in computational fluid dynamics~\cite{Mattsson2003},
and rapid calculation of
slow converging series in electromagnetic problems~\cite{Mosig2002},
it has been particularly overlooked by the
signal processing
community.
Early attempts to employ
it
as a numerical analysis tool
are due to
Boudreaux-Bartels and collaborators
in the context of
the DFT computation~\cite{Boudreaux-Bartels1987}
and
the evaluation of
Fourier coefficients errors calculations~\cite{Boudreaux-Bartels1989}.

The aim of this paper
is
to propose a new fast algorithm for the 8-point DCT
computation
based on
the summation-by-parts formula for
periodic signals~\cite{Graham1989,cintra2012soma}.
The introduced method
is sought to achieve
the theoretical minimal multiplicative complexity
for the exact DCT computation~\cite{Duhamel1987,heideman1988multiplicative}.
Moreover,
to further minimize computational costs,
the proposed algorithm is also sought
to provide a scaled version of the DCT spectrum~\cite{Britanak2007}.
  The proposed algorithm finds application in some important problems,
such as feature detection,
where DC level may not be relevant~\cite{Jain1989,Wang2002}.
Also, it can be applied to scenarios where
input signal is natively accumulated (integrated)~\cite[p.~19]{Oppenheim2009}.
This   situation occurs   in face recognition problems,
where usual algorithms require
data to be integrated~\cite{Elboher2012,Viola2001}.

This paper is organized as follows.
In Section~\ref{sectionmathback},
we furnish
the mathematical background for
the summation-by-parts %
technique and the DCT.
Considering matrix formalism,
we detail the proposed algorithm for the DCT
in Section~\ref{sectionDCT}. In Section~\ref{sectioncomplexity},
the introduced method is assessed in terms
of its computational complexity
and
comparisons with competing algorithms are shown. Section~\ref{sectionconclusion} brings final comments and remarks.

\section{Mathematical Background}
\label{sectionmathback}
\subsection{Summation-by-parts}

The summation-by-parts technique
is the discrete-time equivalent of
the integration-by-parts method~\cite{Graham1989}. Let~$x[n]$ and~$y[n]$ be
two discrete-time signals.
The summation-by-parts prescribes that~\cite{Graham1989,Boudreaux-Bartels1987}:
\begin{align*}
\sum_{n = 0}^{N-1}
x[n] y[n]
=
&
x[N]y[N] - x[0]y[0]
\\
&
-
\sum_{n = 0}^{N-1}
\left(\sum_{i = 0}^{n-1}x[i]\right)\cdot \Delta y[n]
,
\end{align*}
where~$\Delta$ denotes
the forward difference operator given by
$\Delta y[n] \triangleq y[n+1] - y[n]$~\cite{Graham1989}. Above expression can be simplified
with the assumption of the following
additional weak conditions. Admitting
that
the considered signals are periodic
with period~$N$, %
it was established in~\cite{cintra2012soma}
that:
\begin{align}
\label{xy}
\sum_{n = 0}^{N-1}
x[n]\cdot y[n]
=
-
\sum_{n = 0}^{N-1}
\left(\sum_{i = 0}^{n-1}x[i]\right)\cdot \Delta y[n]
.
\end{align}

The above condition is not too restrictive.
Indeed
discrete-time
Fourier analysis
often assume that
the input signals
are
periodic~\cite{Oppenheim2009,Britanak2007,Gonzalez2001}.
In particular,
the DCT can be obtained as
the solution to the harmonic oscillation problem~\cite{Britanak2007}.

The expression
$\sum_{n = 0}^{N-1} x[n]\cdot y[n]$
can be interpreted as
a discrete-time transformation.
Let $x[n]$ be the input signal to be transformed
and
$y[n] = \operatorname{ker}[n, k]$
be
a given discrete transformation kernel
for the $k$th transform-domain component.
Therefore,
we have that:
\begin{align}
\label{Xkgeneral}
X[k]
=
\sum_{n = 0}^{N-1}
x[n]
\cdot
\operatorname{ker}[n, k]
,
\quad
k=0,1,\ldots,N-1
,
\end{align}
where $X[k]$ is the transformed output signal.
Table~\ref{kernels}
summarizes
common transformation kernels.
Therefore,
applying~\eqref{xy}
into~\eqref{Xkgeneral}
yields the following expression for
the transform-domain components:
\begin{align}
\label{Xk}
X[k]
& =
-
\sum_{n = 0}^{N-1}
\left(\sum_{i = 0}^{n}x[i]\right)
\cdot
\Delta \operatorname{ker}[n, k]
\nonumber
\\
& =
-
\sum_{n = 0}^{N-1}
z[n]
\cdot
\Delta \operatorname{ker}[n, k]
,
\quad
k=0, 1,\ldots,N-1,
\end{align}
where
$z[n] = \sum_{i = 0}^{n}x[i]$,
for $n=0,1,\ldots,N-1$.
Comparing~\eqref{Xkgeneral} with~\eqref{Xk},
we notice that
the original transform expression
was
re-written
into an alternative form
where both the input data and the kernel function
were processed. Notice that $z[n]$ is the output of an accumulator system
for input signal~$x[n]$~\cite{Oppenheim2009};
whereas
$\Delta \operatorname{ker}[n, k]$
derives from a forward difference system
for input signal~$\operatorname{ker}[n, k]$~\cite{Oppenheim2009}.
Although the forward difference system is not causal,
this fact poses no difficulty to above formalism. This is because
$\operatorname{ker}[n, k]$ is not a random real-time sequence---but
a deterministic sequence
whose values are known \emph{a priori}~\cite[p.~7]{Hamming1989}.

\begin{table}
\centering
\caption{Common discrete transform kernels}
\label{kernels}
\begin{tabular}[c]{l@{\quad}r@{\quad}}
\toprule
Transform & $\operatorname{ker}[n,k]$
\\
\midrule
DFT~\cite{Oppenheim2009} & $\exp\left(-j\frac{2\pi nk}{N}\right)$ \\
DHT~\cite{Oppenheim2009} & $\operatorname{cas}\left(\frac{2\pi nk}{N}\right)$ \\
DCT~\cite{Britanak2007} & $\cos\left(\frac{\pi(2n+1)k}{2N}\right)$ \\
DST~\cite{Britanak2007} & $\sin\left(\frac{\pi}{N}(k+\frac{1}{2})(n+\frac{1}{2})\right)$ \\
\bottomrule
\end{tabular}
\end{table}

Moreover,
if $x[n]$ possesses null mean,
then the following expression holds true:
\begin{align*}
z[N-1]
=
\sum_{i = 0}^{N-1}
x[i]
=
0
\quad
\text{and}
\quad
X[0]
=
0
.
\end{align*}
For trigonometric transforms,
above condition implies
$X[0]=0$ (null DC~value).
Therefore,
\eqref{Xk}
can be simplified
and written as:
\begin{align}
\label{Xk2}
X[k]
=
-
\sum_{n = 0}^{N-2}
z[n]
\cdot
\Delta
\operatorname{ker}[n, k]
,
\quad
k=1, 2,\ldots,N-1
.
\end{align}
Above summation ranges from $0$ to $N-2$.
This
means that the transformation matrix
linked
to~\eqref{Xk}
has dimension $(N-1) \times (N-1)$.
This fact contrasts with
the original transformation matrix,
which has size $N\times N$.
Thus,
the summation-by-parts
effected a dimension reduction
of transform computation.
As a consequence,
the computational cost of associate algorithms
is expected to be reduced.

\begin{figure}
\centering
\includegraphics{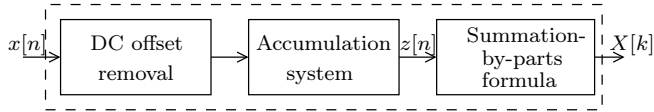}
\caption{Block diagram of the proposed architecture.}
\label{blockdiagram}
\end{figure}

Figure~\ref{blockdiagram}
depicts the
overall diagram
for the
transform computation based on the
summation-by-parts formula,
when
$x[n]$ is assumed to be an arbitrary signal.
Notice that,
if $N$ is a power of two,
both
the DC removal block
and
the accumulation system
are multiplierless operations.

\subsection{Discrete Cosine Transform}

The DCT is a linear transformation
that maps an $N$-point discrete-time signal~$x[n]$
into
another $N$-point discrete-time signal~$X[k]$
according to the following relationship~\cite{Loeffler1989}:
\begin{align}
\label{equation-dct-definition}
X[k]
=
\frac{4}{\sqrt{N}}
\,
\alpha_k
\sum_{n = 0}^{N-1}
x[n]
\cos
\left(
\frac{\pi(2n+1)k}{2N}
\right)
,
\end{align}
where
$k = 0,1,\ldots, N-1$,
$\alpha_0 = 1/\sqrt{2}$,
and
$\alpha_k = 1$, for $k>0$.
The above expression can be given a compact format by means of matrix representation. Indeed, considering
signals~$x[n]$ and~$X[k]$ in
column vector format as
$\mathbf{x} = \begin{bmatrix} x[0] & x[1] & \cdots & x[N-1] \end{bmatrix}^\top$
and
$\mathbf{X} = \begin{bmatrix} X[0] & X[1] & \cdots & X[N-1] \end{bmatrix}^\top$,
we have that:
\begin{align}
\label{mtxXk}
\mathbf{X}
=
\mathbf{C}_N
\cdot
\mathbf{x}
,
\end{align}
where $\mathbf{C}_N$ is the DCT matrix,
whose $(k,n)$-entry is given by
$\frac{4}{\sqrt{N}}
\alpha_k
\cos
\left(
\frac{\pi(2n+1)k}{2N}
\right)
$.
For $N=8$,
we have the following transformation matrix:
\begin{align*}
\mathbf{C}
\triangleq
\mathbf{C}_8
=
\sqrt{2}
\cdot
\left[
\begin{rsmallmatrix}
c_4 & c_4  &     c_4     &  c_4  &     c_4   &    c_4    &   c_4  &     c_4  \\
     c_1   &   c_3  &     c_5  &     c_7    &   -c_7&      -c_5&      -c_3&      -c_1 \\
     c_2   &   c_6   &    -c_6 &     -c_2   &   -c_2&      -c_6  &    c_6  &     c_2  \\
     c_3   &   -c_7  &    -c_1 &     -c_5   &   c_5 &      c_1   &    c_7  &     -c_3 \\
     c_4  &    -c_4  &    -c_4 &     c_4    &   c_4 &      -c_4  &    -c_4 &     c_4  \\
     c_5   &   -c_1  &    c_7  &     c_3    &   -c_3&      -c_7  &    c_1  &     -c_5 \\
     c_6   &   -c_2  &    c_2  &     -c_6  &    -c_6&      c_2   &    -c_2 &     c_6  \\
     c_7   &   -c_5  &    c_3  &     -c_1   &   c_1 &      -c_3   &   c_5  &     -c_7\\
\end{rsmallmatrix}
\right]
,
\end{align*}
where~$c_k = \cos(k\pi/16)$,
for $k=1,2,\ldots,7$.
This particular definition of the DCT
is
adopted
in~\cite{El-Banna2004,JunmingShan2012,Chi-KeungFong2012},
having been considered to derive
the well-known Loeffler DCT algorithm~\cite{Loeffler1989}.
Notice that $\sqrt{2}\cdot c_4 = 1$;
therefore
the~DC component $X[0]$
is evaluated without multiplications~\cite{Loeffler1989}.
In~\cite{heideman1988multiplicative},
Heideman introduces an in-depth mathematical analysis
of the multiplicative complexity of
major discrete transforms.
A result from multiplicative complexity theory
that is relevant to the current work is the following.
If the transform blocklength is a power of two,
$N=2^r$, $r=1,2,\ldots$,
then
the minimum multiplicative complexity $\mu(N)$ of the DCT
has a general formula
given by
$\mu(2^r) = 2^{r+1}-r-2$~\cite[Theorem~6.3, p.~117]{heideman1988multiplicative}.
For $N=8$,
we obtain $\mu(2^3) = 11$.

\section{DCT Computation via Summation-by-parts}
\label{sectionDCT}

In this section,
we apply the summation-by-parts technique to
propose an alternative computation of the
8-point DCT.
Next we analyze the resulting expressions
in order to derive a fast algorithm
by means of matrix factorization.

\subsection{Matrix Formalism}

To facilitate the
development of the sought
DCT fast algorithm,
we re-cast
the summation-by-part formula
into matrix formalism.
But, first,
the
forward difference operator
needs to be adapted to manipulate matrices.
Let $\mathbf{M}$ be a square matrix.
Then
$\Delta\mathbf{M}$ is the matrix
resulted from
cyclically
applying the forward difference operator
to each row of~$\mathbf{M}$.

Therefore,
considering the summation-by-parts formula
shown in~\eqref{Xk2},
we have that~\eqref{mtxXk}
can be written according to:
\begin{align*}
\mathbf{X}
=
\Delta\mathbf{C}
\cdot
\mathbf{z}
,
\end{align*}
where
the $N$-point vector
$\mathbf{X} = \begin{bmatrix} X[0] & X[1] & \cdots & X[N-1] \end{bmatrix}^\top$
represents the DCT spectrum,
$\mathbf{z} = \begin{bmatrix}z[0] & z[1] & \cdots & z[N-1] \end{bmatrix}^\top$
is the accumulated input signal,
and
\begin{align*}
\Delta&\mathbf{C} =
\sqrt{2}\times
\\
&
\left[
\begin{rsmallmatrix}
 0  & 0  & 0  & 0  & 0  & 0  & 0  & 0  \\
 c_3 - c_1  & c_5 - c_3  & c_7 - c_5  & -2c_7  & c_7 - c_5  & c_5 - c_3  & c_3 - c_1  & 2c_1  \\
 c_6 - c_2  & -2c_6  & c_6 - c_2  & 0  & c_2 - c_6  & 2c_6  & c_2 - c_6  & 0  \\
 - c_3 - c_7  & c_7 - c_1  & c_1 - c_5  & 2c_5  & c_1 - c_5  & c_7 - c_1  & - c_3 - c_7  & 2c_3  \\
 -2c_4  & 0  & 2c_4  & 0  & -2c_4  & 0  & 2c_4  & 0  \\
 - c_1 - c_5  & c_1 + c_7  & c_3 - c_7  & -2c_3  & c_3 - c_7  & c_1 + c_7  & - c_1 - c_5  & 2c_5  \\
 - c_2 - c_6  & 2c_2  & - c_2 - c_6  & 0  & c_2 + c_6  & -2c_2  & c_2 + c_6  & 0  \\
 - c_5 - c_7  & c_3 + c_5  & - c_1 - c_3  & 2c_1  & - c_1 - c_3  & c_3 + c_5  & - c_5 - c_7  & 2c_7  \\
\end{rsmallmatrix}
\right]
.
\end{align*}

Considering
the
sum-to-product identities~\cite[p.~72]{MiltonAbramowitz1964}
and symmetry identities~\cite{Graham1989},
the entries of
the matrix
$\Delta\mathbf{C}$
can be given
a multiplicative form,
as shown below:
\begin{align*}
\Delta\mathbf{C} =
2\sqrt{2}
\cdot
\left[
\begin{rsmallmatrix}
0 & 0 & 0 &0 & 0 & 0 & 0 & 0 \\
 s_1 s_{2}  & s_1 s_{4}  & s_1 s_{6}  & s_1  & s_1 s_{6}  & s_1 s_{4}  & s_1 s_{2} & s_{7} \\
 s_2 s_{4}  & s_2  & s_2 s_{4}  & 0  & -s_2 s_{4}  & -s_2  & -s_2 s_{4}  & 0\\
 s_3 s_{6}  & s_3 s_{4}  & -s_3 s_{2}  & -s_3  & -s_3 s_{2}  & s_3 s_{4}  & s_3 s_{6}  & s_{5}\\
 s_4  & 0  & -s_4  & 0  & s_4  & 0  & -s_4  & 0 \\
 s_5 s_{6}  & -s_5 s_{4}  & -s_5 s_{2}  & s_5  & -s_5 s_{2}  & -s_5 s_{4}  & s_5 s_{6}  & s_{3}\\
 s_6 s_{4}  & -s_6  & s_6 s_{4}  & 0  & -s_6 s_{4}  & s_6  & -s_6 s_{4} & 0 \\
 s_7 s_{2}  & -s_7 s_{4}  & s_7 s_{6}  & -s_7  & s_7 s_{6}  & -s_7 s_{4}  & s_7 s_{2}  & s_{1}\\
\end{rsmallmatrix}
\right]
,
\end{align*}
for~$s_k = \sin(k\pi/16)$,
$k=1,2,\ldots,7$.

If
the
input signal
possesses null mean,
then
we have that
$X[0]=0$ and $z[N-1]=0$.
Therefore,
the first row and last column of
$\Delta \mathbf{C}$
can be neglected. It follows that only
the remaining
7$\times$7 submatrix
is necessary for the computation of~$\mathbf{X}$. This particular matrix is given by:

\begin{align*}
\widetilde{\mathbf{C}} =
2\sqrt{2}\cdot\left[
\begin{rsmallmatrix}
 s_1 s_{2}  & s_1 s_{4}  & s_1 s_{6}  & s_1   & s_1 s_{6}  & s_1 s_{4}  & s_1 s_{2} \\
 s_2 s_{4}  & s_2   & s_2 s_{4}  & 0  & -s_2 s_{4}  & -s_2   & -s_2 s_{4}  \\
 s_3 s_{6}  & s_3 s_{4}  & -s_3 s_{2}  & -s_3   & -s_3 s_{2}  & s_3 s_{4}  & s_3 s_{6} \\
 s_4   & 0  & -s_4   & 0  & s_4   & 0  & -s_4   \\
 s_5 s_{6}  & -s_5 s_{4}  & -s_5 s_{2}  & s_5   & -s_5 s_{2}  & -s_5 s_{4}  & s_5 s_{6}  \\
 s_6 s_{4}  & -s_6   & s_6 s_{4}  & 0  & -s_6 s_{4}  & s_6   & -s_6 s_{4} \\
 s_7 s_{2}  & -s_7 s_{4}  & s_7 s_{6}  & -s_7   & s_7 s_{6}  & -s_7 s_{4}  & s_7 s_{2}  \\
\end{rsmallmatrix}
\right].
\end{align*}

Notice that
$\widetilde{\mathbf{C}}$
is sufficient for
the computation of all DCT coefficients---except the DC level.

\subsection{Scaling Matrix}

An examination of matrix~$\widetilde{\mathbf{C}}$
shows
repeated multiplicands along its rows. Thus,
the following
factorization
is obtained:

\begin{align*}
\widetilde{\mathbf{C}}
=
\mathbf{S}
\cdot
\left[
\begin{rsmallmatrix}%
 s_{2}  & s_{4}  & s_{6}  & 1  & s_{6}  & s_{4}  & s_{2}  \\
 s_{4}  & 1  & s_{4}  & 0  & -s_{4}  & -1  & -s_{4}  \\
 s_{6}  & s_{4}  & -s_{2}  & -1  & -s_{2}  & s_{4}  & s_{6}  \\
 1  & 0  & -1  & 0  & 1  & 0  & -1  \\
 s_{6}  & -s_{4}  & -s_{2}  & 1  & -s_{2}  & -s_{4}  & s_{6}  \\
 s_{4}  & -1  & s_{4}  & 0  & -s_{4}  & 1  & -s_{4}  \\
 s_{2}  & -s_{4}  & s_{6}  & -1  & s_{6}  & -s_{4}  & s_{2}  \\
\end{rsmallmatrix}
\right]
,
\end{align*}
where
$\mathbf{S} = 2\sqrt{2} \cdot \operatorname{diag}(s_1, s_2, s_3, s_4, s_5, s_6, s_7)$.
The above expression tells us that
the matrix~$\mathbf{S}$
contributes only with scaling factors
to the actual DCT computation.
When considering applications
that require
only a scaled version of the
DCT---such as
harmonic detection~\cite{Zheng2010}
and
color enhancement~\cite{Mukherjee2008}---the
computational cost of~$\mathbf{S}$
can be disregarded.
Additionally,
in the context of image compression,
diagonal matrices can be
directly absorbed into the quantization matrix;
representing no extra computation~\cite{Bouguezel2008,Dimitrov2005}.

Notice also that
the
scaling factor
$2\sqrt{2}s_4=2$ is a trivial multiplication~\cite{Blahut2010}
that can be implemented
via a simple bit-shifting operation.
Therefore,
the computational cost of the scaling matrix~$\mathbf{S}$
is actually only six---not seven---multiplications.

\subsection{Fast Algorithm}

Now our aim is to provide
a sparse matrix factorization to~$\widetilde{\textbf{C}}$.
Because~$\widetilde{\textbf{C}}$
is a highly symmetrical matrix,
factorization methods based on butterfly structures
can be directly applied~\cite{Oppenheim2009,Britanak2007,Blahut2010}.
Therefore,
we obtain the following
factorization:
\begin{align*}
\widetilde{\textbf{C}} =
\textbf{S} \cdot\textbf{P} \cdot \textbf{M} \cdot \textbf{A},%
\end{align*}
where
\begin{align*}
\mathbf{P} &=
\left[\begin{rsmallmatrix}%
1  & \phantom{-}0  & \phantom{-}0  & \phantom{-}0  & \phantom{-}0  & \phantom{-}0  & \phantom{-}0  \\
 0  & 0  & 1  & 0  & 0  & 0  & 0  \\
 0  & 0  & 0  & 0  & 1  & 0  & 0  \\
 0  & 0  & 0  & 0  & 0  & 0  & 1  \\
 0  & 1  & 0  & 0  & 0  & 0  & 0  \\
 0  & 0  & 0  & 1  & 0  & 0  & 0  \\
 0  & 0  & 0  & 0  & 0  & 1  & 0  \\
\end{rsmallmatrix}\right]
,
\\
\textbf{A} &=
\left[\begin{rsmallmatrix}%
1  & \phantom{-}0  & \phantom{-}0  & \phantom{-}0  & \phantom{-}0  & \phantom{-}0  & \phantom{-}1  \\
 0  & 1  & 0  & 0  & 0  & 1  & 0  \\
 0  & 0  & 1  & 0  & 1  & 0  & 0  \\
 0  & 0  & 0  & 1  & 0  & 0  & 0  \\
 0  & 0  & 1  & 0  & -1  & 0  & 0  \\
 0  & 1  & 0  & 0  & 0  & -1  & 0  \\
 1  & 0  & 0  & 0  & 0  & 0  & -1  \\
\end{rsmallmatrix}\right]
,
\end{align*}
and
\begin{align*}
\mathbf{M} =
\left[\begin{rsmallmatrix}%
s_{2} &s_{4} &s_{6} &1 &0 &0 &\phantom{-}0 \\
s_{6} &s_{4} &-s_{2} &-1 &0 &0 &0 \\
s_{6} &-s_{4} &-s_{2} &1 &0 &0 &0 \\
s_{2} &-s_{4} &s_{6} &-1 &0 &0 &0 \\
0 &0 &0 &0 &s_{4} &1 &s_{4} \\
0 &0 &0 &0 &-1 &0 &1 \\
0 &0 &0 &0 &s_{4} &-1 &s_{4} \\
\end{rsmallmatrix}
\right]
.
\end{align*}

\noindent
Matrix~$\mathbf{P}$ is a simple permutation matrix,
which represents no computational cost.
In terms of hardware,
$\mathbf{P}$ translates into simple wiring.
Matrix~$\mathbf{A}$ is an additive matrix
consisting of the usual butterfly stage
present in decimation-in-frequency algorithms~\cite{Blahut2010}.
The remaining matrix~$\mathbf{M}$
is block-diagonal and still contains
mathematical redundancies
due to its symmetrical nature.
Considering
the matrix factorizations
for fast algorithm design
described in~\cite{Blahut2010},
the following
expression can be obtained:
\begin{align*}
\textbf{M} =
\textbf{M}_1 \cdot \textbf{M}_2 \cdot \textbf{M}_3 \cdot \textbf{M}_4, %
\end{align*}
where
\begin{align*}
\mathbf{M}_1
&=
\left[\begin{rsmallmatrix}%
 1  & \phantom{-}0  & \phantom{-}1  & \phantom{-}0  & \phantom{-}0  & \phantom{-}0  & \phantom{-}0  \\
 0  & 1  & 0  & 1  & 0  & 0  & 0  \\
 0  & 1  & 0  & -1  & 0  & 0  & 0  \\
 1  & 0  & -1  & 0  & 0  & 0  & 0  \\
 0  & 0  & 0  & 0  & 1  & 0  & 0  \\
 0  & 0  & 0  & 0  & 0  & 0  & 1  \\
 0  & 0  & 0  & 0  & 0  & 1  & 0  \\
\end{rsmallmatrix}\right],%
\\
\mathbf{M}_2
&=
\left[\begin{rsmallmatrix}%
 s_2  & \phantom{-}s_6  & \phantom{-}0  & \phantom{-}0  & \phantom{-}0  & \phantom{-}0  & \phantom{-}0  \\
 s_6  & -s_2  & 0  & 0  & 0  & 0  & 0  \\
 0  & 0  & 1  & 1  & 0  & 0  & 0  \\
 0  & 0  & 1  & -1  & 0  & 0  & 0  \\
 0  & 0  & 0  & 0  & 1  & 1  & 0  \\
 0  & 0  & 0  & 0  & 1  & -1  & 0  \\
 0  & 0  & 0  & 0  & 0  & 0  & -1  \\
\end{rsmallmatrix}\right],%
\\
\mathbf{M}_3 &=
\left[\begin{rsmallmatrix}%
1  & \phantom{-}0  & \phantom{-}0  & \phantom{-}0  & \phantom{-}0  & \phantom{-}0  & \phantom{-}0  \\
 0  & 0  & 1  & 0  & 0  & 0  & 0  \\
 0  & s_4  & 0  & 0  & 0  & 0  & 0  \\
 0  & 0  & 0  & 1  & 0  & 0  & 0  \\
 0  & 0  & 0  & 0  & s_4  & 0  & 0  \\
 0  & 0  & 0  & 0  & 0  & 1  & 0  \\
 0  & 0  & 0  & 0  & 0  & 0  & 1  \\
\end{rsmallmatrix}
\right],
\\
\mathbf{M}_4 &=
\left[\begin{rsmallmatrix}%
 1  & \phantom{-}0  & \phantom{-}0  & \phantom{-}0  & \phantom{-}0  & \phantom{-}0  & \phantom{-}0  \\
 0  & 1  & 0  & 0  & 0  & 0  & 0  \\
 0  & 0  & 1  & 0  & 0  & 0  & 0  \\
 0  & 0  & 0  & 1  & 0  & 0  & 0  \\
 0  & 0  & 0  & 0  & 1  & 0  & 1  \\
 0  & 0  & 0  & 0  & 0  & 1  & 0  \\
 0  & 0  & 0  & 0  & 1  & 0  & -1  \\
\end{rsmallmatrix}\right]
.
\end{align*}

The block-diagonal matrix~$\mathbf{M}_2$
contains
a
rotation block
$\left[\begin{rsmallmatrix}s_2 & s_6 \\ s_6 & -s_2\end{rsmallmatrix}\right]$,
which can be further decomposed~\cite{Britanak2007}.
Thus,
we have:
\begin{align*}
\mathbf{M}_2 =
\mathbf{R}_1
\cdot
\mathbf{R}_2
\cdot
\mathbf{R}_3
,
\end{align*}
where
\begin{align*}
\mathbf{R}_1 &=
\left[\begin{rsmallmatrix}%
 0  & s_{6} - s_{2}  & 1  & 0  & 0  & 0  & 0  & 0  \\
 s_{2} + s_{6}  & 0  & -1  & 0  & 0  & 0  & 0  & 0  \\
 0  & 0  & 0  & 1  & 0  & 0  & 0  & 0  \\
 0  & 0  & 0  & 0  & 1  & 0  & 0  & 0  \\
 0  & 0  & 0  & 0  & 0  & 1  & 0  & 0  \\
 0  & 0  & 0  & 0  & 0  & 0  & 1  & 0  \\
 0  & 0  & 0  & 0  & 0  & 0  & 0  & 1  \\
\end{rsmallmatrix}\right]
,
\\
\mathbf{R}_3 &=
\left[\begin{rsmallmatrix}%
 1  & \phantom{-}0  & \phantom{-}0  & \phantom{-}0  & \phantom{-}0  & \phantom{-}0  & \phantom{-}0  \\
 0  & 1  & 0  & 0  & 0  & 0  & 0  \\
 1  & 1  & 0  & 0  & 0  & 0  & 0  \\
 0  & 0  & 1  & 1  & 0  & 0  & 0  \\
 0  & 0  & 1  & -1  & 0  & 0  & 0  \\
 0  & 0  & 0  & 0  & 1  & 1  & 0  \\
 0  & 0  & 0  & 0  & 1  & -1  & 0  \\
 0  & 0  & 0  & 0  & 0  & 0  & -1  \\
\end{rsmallmatrix}
\right]
,
\end{align*}
and~$\mathbf{R}_2 = \operatorname{diag}(1, 1, s_2, 1, 1, 1, 1, 1)$.

By means of usual trigonometric manipulations,
we notice that
the required multiplicands in~$\mathbf{R}_1$
satisfy:
$s_2 + s_6 = \sqrt{2}\,c_2$
and
$s_6 - s_2 = \sqrt{2}\,s_2$. Finally,
the   complete  sparse matrix factorization is given by:
\begin{align*}
\widetilde{\mathbf{C}} =
\mathbf{S}\cdot \mathbf{P}\cdot \mathbf{M}_1\cdot \mathbf{R}_1\cdot \mathbf{R}_2\cdot \mathbf{R}_3\cdot \mathbf{M}_3\cdot \mathbf{M}_4\cdot \mathbf{A}. %
\end{align*}

\begin{figure}
\centering
{\includegraphics{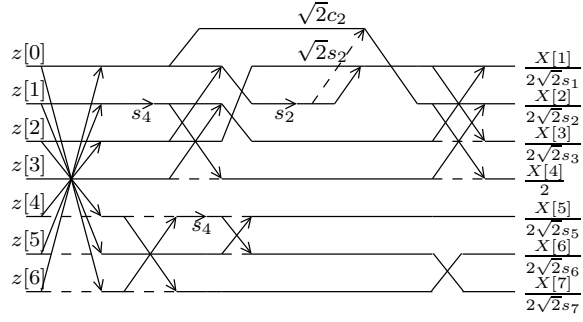}}
\caption{%
SFG for the proposed algorithm.
Dashed  lines represents multiplication by~$-1$.}
\label{flowgraph}
\end{figure}

The signal flow graph (SFG) for the
proposed algorithm is shown in Figure~\ref{flowgraph}.

\begin{figure}
\centering
{\includegraphics{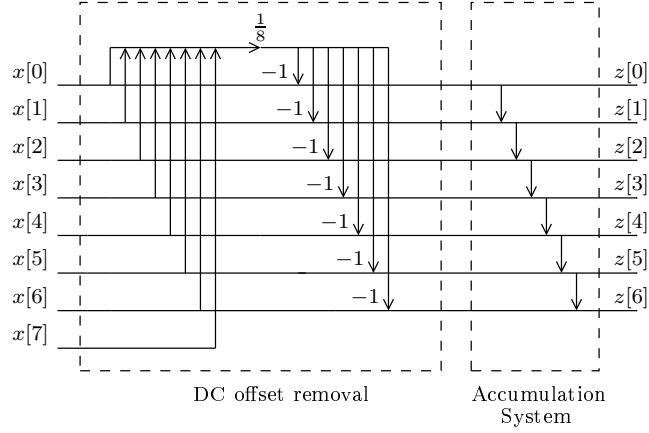}}
\caption{%
SFG for pre-processing stage consisting
of
the DC component removal coupled with accumulation system.}
\label{DCremovalAndAccumulation}
\end{figure}

\section{Computational Complexity Assessment and Comparison}
\label{sectioncomplexity}

In this section,
we assess the computational complexity of
the proposed algorithm
as measured by its arithmetic cost.
For such,
we evaluate the additive and multiplicative count
required by the introduced method.

\subsection{Results}

We distinguish the following scenarios
for the input data:
(i)~arbitrary signal;
(ii)~null mean signal;
(iii)~accumulated signal;
and
(iv)~null mean and accumulated signal.
Scenario~(i)
is the most general case.
Thus,
there is less redundancy to be exploited
aiming at the minimization of computational cost.
Scenario~(i) is commonly found
in image compression applications~\cite{Britanak2007,Gonzalez2001}.
Scenario~(ii)
is pertinent in the context of
feature detection, for instance,
where DC level may not be relevant~\cite{Jain1989,Wang2002}.
Scenario~(iii)
represents the case where
the input signal is natively accumulated (integrated).
This situation occurs in face recognition problems,
where usual algorithms require
data to be integrated~\cite{Elboher2012,Viola2001}.
Scenario~(iv) is a combination of above last two cases.
Notice that
Scenario~(ii) does not require
the DC offset removal block (cf.~Figure~\ref{blockdiagram});
Scenario~(iii) does not require the accumulation system
but needs a DC removal block;
and
Scenario~(iv) requires neither the DC offset removal block nor the accumulation system.

Table~\ref{algcomparisons}
compares the proposed method
with several prominent techniques for
the 8-point DCT evaluation
under all discussed scenarios.
The details of the arithmetic complexity
assessment of each considered technique
is found in~\cite{Britanak2007}.
For each scenario,
we emphasized in bold the best results.
Because
Arai method~\cite{Arai1988} and the proposed algorithm
are scaled DCT methods,
we show the multiplicative complexity of
the non-scaled computation as well.
In parenthesis,
when applicable,
we furnish
the full multiplicative complexity of the scaled methods,
when the scaling factors are considered.
The multiplicative complexity for Loeffler, Lee, and Chen DCT algorithms
are limited to the non-scaled case,
because these methods do not admit scaled computation.
The proposed algorithm
could outperform
\emph{all} competing
method under Scenarios~(ii), (iii), and~(iv).
Although
the proposed method
requires five multiplications
to compute the scaled
DCT---the same as the Arai method---it demands fewer scaling multiplications.
Scenario~(i)
is not ideally suitable
for the proposed algorithm
because
it benefits of
the DC level removal and
accumulated input signal.
However,
even in this case,
the introduced algorithm
could achieve
the theoretical minimum of multiplicative complexity.

\begin{table}%
\small
\centering
\caption{Comparison for non-trivial products and additions for $8$-point DCT algorithms}
\label{algcomparisons}
\begin{tabular}[c]{l@{ }c@{\quad}c@{\quad}c@{\quad}c@{}}
\toprule
Algorithm &
Scaled? &
Scenarios &
Mult. &
Additions \\
\midrule
\multirow{4}{*}{Loeffler~\cite{Loeffler1989}} & \multirow{4}{*}{No}  &  (i) & $\mathbf{11}$ & $\mathbf{29}$ \\
&   & (ii) & $11$ & $26$\\
&   & (iii) &$11$ & $36$\\
&   & (iv) & $11$& $33$\\ \midrule
\multirow{4}{*}{Lee~\cite{Lee1984}} & \multirow{4}{*}{No} &  (i) & $12$ & $29$\\
&   &  (ii) &$11$ & $26$\\
&   &  (iii) & $12$&$36$ \\
&   &  (iv) & $11$& $33$ \\ \midrule
\multirow{4}{*}{Chen~\emph{et al.}~\cite{Wen-HsiengChen2003}} & \multirow{4}{*}{No}  &  (i) & $13$ & $26$ \\
&   &  (ii) & $12$& $23$ \\
&   &  (iii) & $13$ & $33$ \\
&   &  (iv) & $12$& $30$ \\
\midrule
\multirow{4}{*}{Arai~\emph{et al.}~\cite{Arai1988}} & \multirow{4}{*}{Yes} &
       (i) &  $\mathbf{5}$ ($13$)  & $\mathbf{28}$ \\
&   &  (ii)  &  $5$ ($12$)  & $25$ \\
&   &  (iii) &  $5$ ($13$)  & $35$ \\
&   &  (iv)  &  $5$ ($12$)  & $32$ \\
\midrule
\multirow{4}{*}{Proposed} & \multirow{4}{*}{Yes} &
       (i)   &  $5$ ($11$) & $39$\\
&   &  (ii)  &  $\mathbf{5}$ ($\mathbf{11}$) & $\mathbf{25}$ \\
&   &  (iii) &  $\mathbf{5}$ ($\mathbf{11}$) & $\mathbf{30}$ \\
&   &  (iv)  &  $\mathbf{5}$ ($\mathbf{11}$) & $\mathbf{19}$ \\
\midrule
\end{tabular}
\end{table}

\subsection{Discussion}

For Scenario~(i),
the DC offset removal block requires seven additions
to compute the mean value and
then
seven subtractions to remove the DC level from
input samples
$x[0], x[1], \ldots, x[6]$.
Thus, a total of 14~additions are necessary.
The accumulation system
requires
a total of six additions (cf.~\eqref{Xk2}).
Thus,
the DC removal block combined with accumulation system
requires a total of~20~additions.
Figure~\ref{DCremovalAndAccumulation}
details the inner structure of
each of the above-mentioned blocks.
Multiplicative costs are concentrated in
matrices~$\mathbf{S}$,
$\mathbf{R}_1$,
$\mathbf{R}_2$,
and~$\mathbf{M}_3$.
They account for $11$~non-trivial
multiplications,
which is the minimum multiplicative complexity
for the $8$-point DCT~\cite{Duhamel1987,heideman1988multiplicative}.
Matrices~$\mathbf{M}_1$,
$\mathbf{R}_1$,
$\mathbf{R}_3$,
$\mathbf{M}_4$,
and~$\mathbf{A}$
contribute with $19$~additions.
Thus,
under Scenario~(i),
the proposed algorithm requires a total of~39~additions.
As we demonstrate below,
this is the only scenario where the proposed method
cannot outperform the most relevant method for the 8-point DCT
computation.
Nevertheless,
even in this scenario,
the proposed method attained the
theoretical minimal multiplicative complexity,
which is a relevant figure of merit~\cite{Blahut2010}.
In comparison with competing methods,
the extra required computation consists of additions only.

\begin{figure}
\centering
\subfigure[\label{DCremovalFromAccumulated}]%
{\includegraphics{DCremovalFromAccumulated}}
\qquad
\subfigure[\label{differentiator}]%
{\includegraphics{differentiator}}
\caption{%
SFG of (a)~the DC removal block and (b)~the finite difference operator,
both for accumulated input signal.}
\label{SFGsAux}
\end{figure}

Under Scenario~(ii),
the proposed algorithm does not demand the DC removal block
but only
one instantiation of an accumulation system
(Figure~\ref{DCremovalAndAccumulation}, rightmost subblock).
Therefore,
a total of 13~additions in savings
is attained
when compared to Scenario~(i).
Consequently,
considering Scenario~(ii),
the proposed algorithm requires only~$25$ additions.
Usual methods presented in literature can be adapted for Scenario~(ii).
However,
when compared to the costs under Scenario~(i),
such methods could save only three additions,
instead of seven additions
for the computation of the DC~level.
This is particularly true
for both Loeffler~\cite{Loeffler1989} and Lee~\cite{Lee1984} algorithms
which demand the same number of multiplications
as the proposed method when exact spectrum is required.
Additionally,
the proposed algorithm
can furnish a scaled version of the DCT spectrum
at the same performance required by the Arai algorithm~\cite{Arai1988}.

Under Scenario~(iii),
the proposed method does not demand
the accumulation system,
but still requires the DC removal block.
In this case,
the DC removal block must be adapted to process
accumulated input signals.
Figure~\ref{DCremovalFromAccumulated}
shows the modified DC removal block,
which
requires~10~additions
to generate the sought sequence~$z[n]$.
In contrast,
traditional methods for DCT computation
when faced with accumulated signals
have to pre-process the input
data through
a difference system.
Figure~\ref{differentiator} shows the SFG for a difference system,
which is completely \emph{eliminated} in the proposed algorithm.

Scenario~(iv) is a combination of Scenarios~(ii) and~(iii);
being
the most appropriate scenario
for the proposed method.
In this case,
the proposed algorithm does not require any pre-processing stage,
being the input data directly applicable.
Thus,
the proposed method is limited
the operations described in the architecture shown
in Figure~\ref{flowgraph}.
Under this scenario,
usual methods requires
a pre-processing stage consisting of a difference system
(Figure~\ref{differentiator}),
which demands seven extra additions.

The proposed algorithm can be extended for different sequence sizes,
which can benefit multiple-blocklength encoding methods
such as HEVC~\cite{Pourazad2012,Britanak2007}.
Such extension
can be derived from~\eqref{Xk}
assuming an
arbitrary length~$N$.
Considering the DCT kernel
(Table~\ref{kernels})
and applying the forward difference operator,
we obtain:
\begin{align*}
\Delta
\operatorname{ker}[n, k]
=
2\alpha_k
\sin\left( \frac{k\pi}{2N}\right)
\sin\left( \frac{k\pi(n+1)}{N} \right)
,
\end{align*}
where
$\alpha_k$
is given as in~\eqref{equation-dct-definition}
and~$n, k = 1, 2, \ldots, N-1$.
Despite
the similarity with the DST~kernel,
the above expression contains
a post-multiplicative
factor~$2\alpha_k\sin\left( k\pi/2N \right)$
that
can be separated
into
a
diagonal matrix.
Thus the design of the associate fast algorithm
becomes
equivalent
to the problem
of factorizing
an
$(N-1)\times(N-1)$ matrix
with
entries
given by~$\sin\left( k\pi(n+1)/N \right)$,
$n, k = 1, 2, \ldots, N-1$.
Such factorization can be accomplished
by means of the techniques detailed in~\cite{Blahut2010}
and the approach described in this paper,
where the 8-point DCT is a showcase.

The proposed 1D~DCT algorithm can be readily extended to the
two dimensional (2D) case,
thus being
suitable
for
image processing applications.
Indeed,
because of DCT kernel separability,
the 2D~DCT can be computed by successive calls
of the 1D~DCT.
This can be achieved in a two-step procedure:
(i)~computation of the 1D~DCT of each row of a given input image;
and
(ii)~computation of the 1D~DCT of each column of
the image derived from step~(i).
The resulting image is the 2D~DCT of the original image.
Therefore,
in principle,
any 1D~DCT algorithm can be immediately
extended to the 2D~case
without any additional modification~\cite{Britanak2007}.

Regarding the DCT computation over 2D domain, different image representation schemes can be employed in order to reduce the overall execution time.
The works on~\cite{Papakostas2009,Papakostas2008,Papakostas2008a,Papakostas2007}  consider a different image representation based on slice intensity representation (ISR).
Although this representation can be useful for some scenarios such as pattern recongnition~\cite{Hassaballah2016,Sayyouri2012,Hmimid2012}, we attain to the usual image representation adopted by the signal processing community~\cite{Blahut2010,Britanak2007,Ahmed1975}.

\color{black}

\section{Remarks and Conclusion}
\label{sectionconclusion}

In this paper,
we proposed
a new method for the 8-point DCT computation
based on the summation-by-parts formula.
A matrix formalism was furnished and
its arithmetic complexity was assessed.

The proposed method attained the theoretical multiplicative complexity
lower bound for the computation of the \emph{exact} DCT
as detailed in~\cite{heideman1988multiplicative}.
The theoretical minimum is consists of 11~multiplications.
\emph{Per se},
achieving the minimum of 11~multiplications is
not a trivial task.
Not many DCT algorithms are capable of such,
being the Loeffler DCT the most popular method in literature
to achieve the minimum~\cite{Loeffler1989}.
Moreover,
the introduced method could also compute the scaled DCT with
only \emph{five} multiplications,
matching the well-known Arai DCT~\cite{Arai1988}.
Thus the proposed method can simultaneously match
both Loeffler and Arai DCT
in terms of multiplicative complexity.
Not only it could attain the multiplicative complexity
minimum,
but it could also
outperform several competing methods
in three different scenarios
when we also consider the additive complexity as a secondary
comparison metric.
In fact,
it
outperformed
several popular DCT algorithms,
when
the input signal
is assumed
to possess null mean
or/and
it is an integrated (accumulated) signal.

The proposed design
can be understood as a fundamental mathematical block
to be considered for software and hardware realizations.
Additionally,
the field of integer and approximate transforms
can benefit of the proposed scheme~\cite{Bouguezel2013,Bouguezel2011}.
Indeed,
the design of extremely low-complexity
approximation methods
relies on exact algorithms.
Therefore,
exact, algebraically precise methods
can be a useful resource;
even when an approximation could be enough~\cite{tablada2015class}.
Particularly,
the proposed method is an alternative to Arai algorithm~\cite{Arai1988}
for selected scenarios.
The suggested method is suitable in the
context of
feature detection,
where DC level may not be relevant~\cite{Wang2002}
as well as in situations where
traditional algorithms require
data to be integrated,
such as in face recognition problems~\cite{Elboher2012,Viola2001}.
In future works,
we aim at applying
the summation-by-parts formula
to
different transforms
at
various blocklengths.

\section*{Acknowledgements}
This work was partially funded by CNPq, CAPES, and FACEPE.

{\small
\singlespacing
\bibliographystyle{ieeetr}
\bibliography{bibcleanoutput}
}

\end{document}